# Lessons Learned from Structural Design and Vibration Testing of 50-kg Microsatellites Deployed from the International Space Station

By Yuji Sakamoto[1], Junichi Kurihara[2], Shinya Fujita[3], Yuji Sato[3], Toshinori Kuwahara[3]

[1] *Green Goals Initiative, Tohoku University, Sendai, Japan*
[2] *Department of Earth and Planetary Sciences, Hokkaido University, Sapporo, Japan*
[3] *Department of Aerospace Engineering, Tohoku University, Sendai, Japan*

Hokkaido University and Tohoku University have been developing and operating a constellation of 50-cm-class microsatellites for Earth observation. DIWATA-1, launched in 2016, was deployed into a circular orbit at an altitude of approximately 400 km from the International Space Station (ISS). For the subsequent satellite developed in 2021, the structural design and vibration test campaign were optimized to meet a strict one-year development schedule. This paper summarizes how the structural design of the previous satellite was reviewed and updated, and how the vibration test was successfully completed in a single trial to minimize schedule and technical risks. These lessons learned provide valuable insights, as there are only a limited number of reported cases of 50-kg-class microsatellites deployed from the ISS.

## 1. Introduction

Hokkaido University and Tohoku University are developing and operating a group of 50-cm-class microsatellites for Earth observation. The main payloads are the High Precision Telescope (HPT) and the Spaceborne Multispectral Imager (SMI), which have been demonstrated in orbit[1)2)].

In 2016, the 50-cm-class satellite DIWATA-1 (D1) was jointly developed with the Philippine government and was deployed into a 400-km-altitude circular orbit from the International Space Station (ISS). It also achieved significant results as a capacity-building program for training young engineers in the Philippines[3]. D1 demonstrated onboard equipment and software originally developed for the RISESAT satellite (2019). It contributed to the evaluation of attitude determination and control capabilities and to improvements in accuracy using operational data. In particular, the HPT requires target tracking within a 0.1-deg error, and updates to the attitude evaluation simulator installed in the development laboratory were achieved based on actual operational experience[4)-6)].

Satellites launched into a 400-km-altitude orbit are highly suitable for demonstration and training purposes. Such satellites re-enter the atmosphere within a relatively short period, reducing concerns regarding space debris. In the case of CubeSats weighing less than 3 kg, there are many examples in which operations are completed within one year; however, for D1, which had a mass of 52.4 kg, operations continued for a total of 3 years and 11 months. Compared with satellites at an altitude of 500 km, the radiation environment affecting onboard equipment is less severe, resulting in a lower risk of radiation-induced damage. More than three years of operation of a practical remote-sensing satellite provided valuable experience for young engineers and scientists.

In 2021, the latest satellite, RISING-4 (hereafter referred to as R4), which is a development codename, was deployed into orbit from the ISS, as shown in Fig. 1. R4 inherits and improves upon the structural design of D1. In this paper, the improvement methods applied during the development of R4 are described, utilizing the knowledge obtained through the design and test evaluations of D1. As a result, the flight-model (FM) vibration tests of R4 were completed without any problems. These lessons learned are rare and valuable, given the limited number of reported cases of 50-cm-class satellites deployed from the ISS.

After structural components and electronic equipment were procured, the flight-model assembly and evaluation tests were carried out during the final 1.5 months of the development period. The vertical vibration test was predicted to be the most challenging task in the final phase. The satellite width of 55 × 55 cm is significantly larger than the 30-cm diameter of the vibration test table. Therefore, it was necessary to newly design a lightweight and highly rigid jig and to accurately predict the natural frequency of the combined satellite-jig system. Vibration tests were preliminarily conducted at three test sites using iron dummy-mass plates. Finally, the horizontal and vertical vibration tests of R4 were carried out separately at different test sites, and the test results satisfied all requirements without any issues.

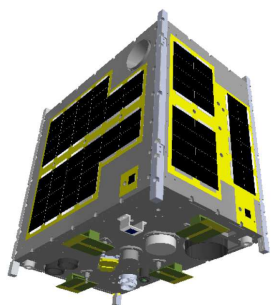

Fig. 1. Appearance of a 50-cm-class microsatellite deployed from the International Space Station

Table 1. Specification of structural subsystem

| Satellite Name | DIWATA-1 (D1) | | RISING-4 (R4) | |
|---|---|---|---|---|
| Dimension (mm) | 550 x 350 x 550 | | 550 x 350 x 550 | |
| Total mass (kg) | **52.4** | | **50.5** | |
| incl. (kg) | | | | |
| Mission payloads | 9.0 | (17.2 %) | 8.4 | (16.5 %) |
| ADCS | 9.7 | (18.4 %) | 9.9 | (19.6 %) |
| Power | 6.1 | (11.6 %) | 6.1 | (12.0 %) |
| Communication | 2.5 | (4.8 %) | 1.8 | (3.5 %) |
| C&DH | 1.9 | (3.6 %) | 2.1 | (4.1 %) |
| Structure | 20.9 | (39.8 %) | 18.7 | (37.0 %) |
| Harness | 2.4 | (4.6 %) | 3.7 | (7.2 %) |

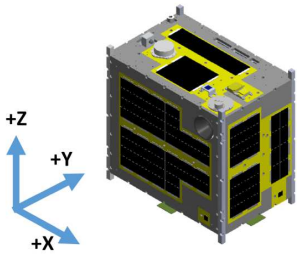

Fig. 2. Definition of structural coordinate frame: origin at the center of the −Z side rail edge plane

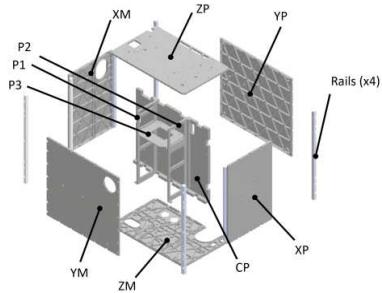

Fig. 3. Deployment view of R4 main structural panels and frames

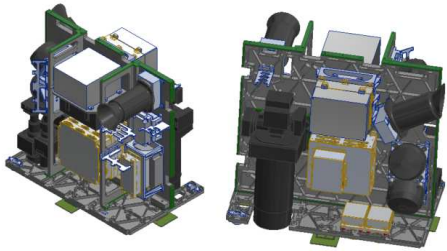

Fig. 4. Internal appearance of R4 satellite: YM side (left) and YP side (right)

## 2. Structure Design and Test Evaluation

### 2.1. Overview of main structural parts

Table 1 shows a comparison of the structural masses of the D1 and R4 satellites. D1 has a total mass of 52.4 kg, while R4 has a mass of 50.5 kg. The mass of the structural parts was reduced by 2.2 kg; however, the mass of the wire harnesses increased by 1.3 kg. For the other subsystems, mass changes of less than +200 g or reductions of 600–700 g were achieved compared with D1.

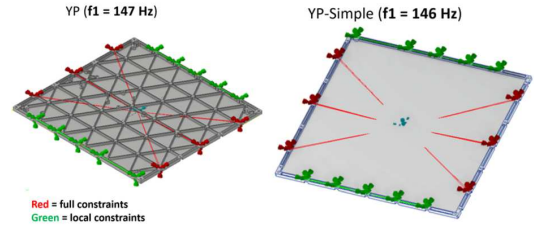

Fig. 5. Normal mode analysis of YP original panel (left) and YP simplified model (right)

Table 2. Mass properties of real and simplified panels in the analysis model

| Part | Real | | | | | Simple | |
|---|---|---|---|---|---|---|---|
| | t of skin (mm) | t of rib (mm) | str. mass (g) | compo. mass (g) | total mass (g) | t of skin (mm) | density (g/cm^3) |
| YP | 1.0 | 1.5 | **1329** | 415 | **1745** | 4.2 | **1.61** |
| YM | 1.0 | 1.5 | **1324** | 693 | **2017** | 4.1 | **1.95** |
| XP | 1.0 | 1.5 | **1605** | 555 | **2159** | 5.2 | **2.06** |
| XM | 1.0 | 1.5 | **1563** | 208 | **1771** | 5.0 | **1.81** |
| ZP | 1.0 | 1.5 | **1166** | 1248 | **2414** | 3.7 | **3.52** |
| ZM | 1.0 | 2.0 | **1230** | 1866 | **3096** | 4.0 | **4.66** |
| CP | 1.6 | 2.5 | **2934** | 25197 | **28131** | 6.0 | **17.71** |
| P1 | | | **794** | 0 | **794** | | **2.80** |
| P2 | | | **804** | 0 | **804** | | **2.80** |
| P3 | | | **383** | 4081 | **4464** | | **32.61** |
| Rail x4 | | | **1844** | 0 | **1844** | | **2.80** |
| | | | | total | **49238** | | |

Figure 2 shows the definition of the structural coordinate system. The longitudinal direction of the rails corresponds to the Z-axis. The satellite is deployed into orbit from the container in the +Z direction. Before deployment, each rail is compressed by spring force along the Z-axis direction.

Figure 3 shows the development view of the main structural panels. Each panel has an aluminum grid design, in which triangular sections are cut out from A7075-T7351 plate material, while rib lines are retained to maintain stiffness. The panels and frames are fastened with stainless-steel or titanium bolts. The ZP, ZM, XP, XM, YP, and YM panels are located on the outer side, the CP panel is located on the inner side, and the P1, P2, and P3 frames support the panels. Figure 4 shows an image of the installed internal equipment.

### 2.2. Structural analysis

For the R4 satellite, it is required to demonstrate through analysis that the natural frequency (normal mode) of the entire satellite is 60 Hz or higher, and that no deformation or damage occurs under a static acceleration of 7 G applied along each axis. In addition, it is necessary to apply a specified-level random vibration test to measure the natural frequency and to confirm that the structure and mounted components, particularly fragile glass parts, are not detached or damaged.

Autodesk® Inventor® 2020 and Inventor Nastran® were used for this analysis. Since the three-dimensional model data of each structural component are directly used to generate

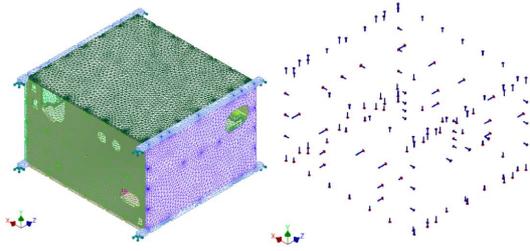

Fig. 6. Mesh model using solid elements (left) and bolt connectors (right)

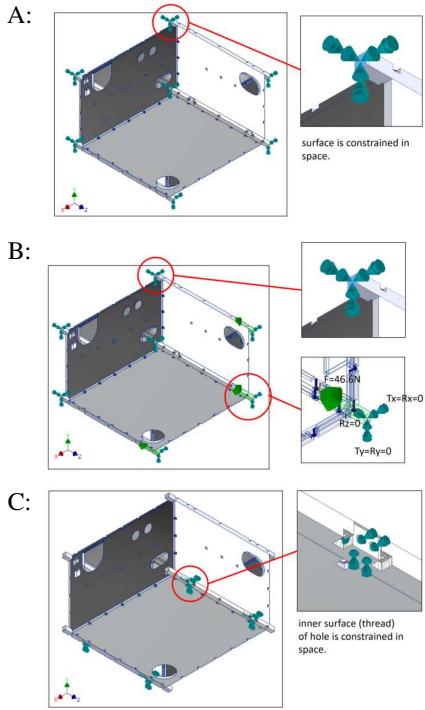

Fig. 7. Constraint conditions: normal mode analysis at launch (A), static load analysis at launch (A), and normal mode analysis during vibration testing (C)

manufacturing drawings, the actual shapes are modeled in the CAD environment with high accuracy. When analyzing a single structural component, the analysis can be rapidly performed using solid elements, and the calculation is completed within a relatively short time, typically less than one minute or a few minutes. In contrast, analyzing the entire assembly as solid elements with a complex grid design requires a large amount of computational time and is inefficient, particularly during the design phase. Therefore, each component is replaced with a simplified shape having equivalent stiffness. The mass of onboard equipment is distributed to the panel masses, and the density of each panel is adjusted to a dummy value in order to reproduce the natural frequencies.

An example of the simplification method is shown using the YP panel in Fig. 5. First, the displacement and rotation at six locations (red) at the outer bolt holes are fully constrained in space, while ten locations (green) are mutually constrained by rigid bars and left unconstrained in space. This constraint condition provides a value close to the local natural frequency of the YP panel obtained during the FM vibration test of D1.

Table 3. Results of normal mode analysis under constraint conditions A and C

| | Frequency (Hz) | | Effective Mass (kg)* | | |
| --- | --- | --- | --- | --- | --- |
| Mode | Const-A | Const-C | X | Y | Z |
| 1 | **96.1** | **97.2** | | 23.3 | |
| 2 | **99.9** | **99.3** | | | 6.9 |
| 3 | **106.6** | **107.5** | 46.4 | | |
| 8 | -- | 164.7 | | | 9.6 |
| 9 | -- | 177.3 | | | 26.2 |
| 10 | -- | 192.7 | 5.7 | | |

* case of Const-B, only >5.0 kg

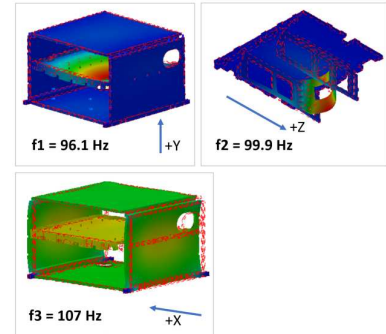

Fig. 8. Normal modes with the lowest frequencies along each vibration axis

The simplified model, referred to as YP-Simple, has a flat-plate region obtained by simplifying the original grid design, while retaining the same design around the side bolt holes. The residual thickness of the central flat region is determined to provide equivalent rigidity and natural frequency.

Table 2 lists the densities of each panel used in the analysis. For the YP panel, in the actual grid structure, the residual thickness of the thin plate is 1.0 mm, the rib width is 1.5 mm, and the mass of the panel alone is 1329 g. When 415 g of mounted components are added, the total mass becomes 1745 g. In the simplified YP-Simple model, the residual thickness of the central region is 4.2 mm, and the density is calculated as 1.61 g/cm$^3$ in order to reproduce the same combined mass of the panel and mounted equipment. In particular, many instruments are fastened to the CP panel, resulting in a density of 17.71 g/cm$^3$, which is larger than that of the other panels. The total mass of all components in the analysis model is 49.2 kg, corresponding to 97.4% of the final FM mass of 50.5 kg.

Figure 6 shows the analysis model after mesh generation. The model consists of 370,678 nodes and 191,423 elements generated using the automatic meshing function of the software. Surface contacts are not defined. Bolt connections between panels are modeled using bolt connectors, which are computationally converted into beam elements. The washer components are not modeled as individual parts; instead, equivalent solid elements with the same diameter and height are modeled on the A7075 panel side, although the original washer material is stainless steel.

M4-size stainless steel bolts (SUS304) are used in the analysis, although some of them are replaced with titanium bolts in the actual structure. Locking helicoils are applied to all bolt threads in the panels, with a standard size of 2.0D and an 8-mm female thread length. Because a protrusion exists in the middle of the thread to increase friction with the bolt, the bolt length is determined to exceed the helicoil length by more than 4 mm.

The constraint conditions differ between the natural frequency analysis and the static load analysis under launch conditions, as shown in Fig. 7. In constraint condition A, the rail end face on the ± Z side is fully restrained in both displacement and rotation. In constraint condition B, since a compressive force of 46.6 N is applied to each rail, displacement in the Z direction is released on only one side. In addition, stresses acting on the panels are calculated by applying a static acceleration of 7 G along each axis.

The satellite developer determines the policy for fixing the satellite during vibration testing. For both the D1 and R4 satellites, the two rails on the YM side are fixed to the jig plate using a total of eight bolts, which corresponds to constraint condition C.

Table 3 shows the results of the natural frequency analysis under constraint conditions A and C. The normal modes for constraint A are graphically shown in Fig. 8. For the first to third modes, the frequency difference between conditions A and C is less than 1.2%. This indicates that the vibration test results obtained under constraint condition C can be regarded as equivalent to the in-orbit natural frequencies under constraint condition A. The natural frequencies are 96.1 Hz (Y axis), 99.9 Hz (Z axis), and 106.6 Hz (X axis). The direction of each mode is estimated based on the effective mass.

Table 4 and Fig. 9 show the results of the static load analysis under constraint condition B. By applying an acceleration of 7.0 G independently along each axis, it was confirmed that the maximum von Mises stress, $S_{max}$, of the solid elements satisfies the safety margins $MS_{ty}$ and $MS_{tu}$ against the yield stress ($F_{ty}$ = 385 MPa) and ultimate stress ($F_{tu}$ = 460 MPa), respectively, for the A7075-T7351 material.

$$MS_{ty} = \frac{F_{ty}}{S_{max}*1.5} - 1 = \frac{385\ MPa}{S_{max}*1.5} - 1 > 0 \quad (1)$$

$$MS_{tu} = \frac{F_{tu}}{S_{max}*2} - 1 = \frac{460\ MPa}{S_{max}*2} - 1 > 0 \quad (2)$$

The maximum stress $S_{max}$ occurs at the bolt washer location where the rail is fastened to the panel for all axis accelerations. The maximum stresses are 68.8 MPa along the X axis, 37.8 MPa along the Y axis, and 41.0 MPa along the Z axis. The minimum safety margins $MS_{ty}$ and $MS_{tu}$ are 2.73 and 2.34, respectively, for the X-axis case. In addition, a requirement of $S_{max} / F_{tu} < 30\%$ is imposed for 50-cm-class satellites deployed from the ISS, and the maximum value obtained in this analysis is 15.0% along the X axis.

**2.3. Increase in stiffness by the number of fastening bolts**

Compared with D1, R4 increases the number of fastening bolts used for the main structural components. While a total of 84 bolts were used in D1, 112 bolts are used in R4, representing an increase of 28 bolts. Around the rail sections of R4 shown in Fig. 10, the number of red-marked bolts is increased from 3 to 7, and the number of orange-marked bolts is increased from 5 to 7 for each rail.

Figure 11 shows the results of the natural frequency analysis when a fastener bolt configuration equivalent to that

Table 4. Margins of safety in static load analysis under constraint condition B

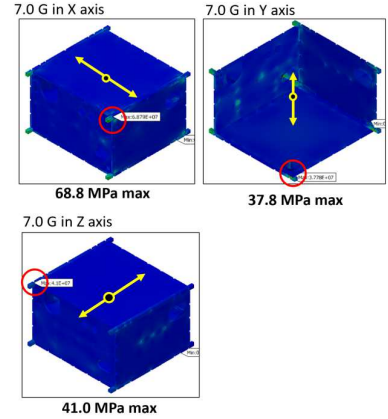

| Condition | Max Stress (Smax, MPa) | Fty | Ftu | MS (yield, FS=1.5) | MS (ultimate, FS=2) | Smax/Ftu <30[%] |
|---|---|---|---|---|---|---|
| X, 7.0G | 68.8 | 385 | 460 | 2.73 | 2.34 | 15.0 |
| Y, 7.0G | 37.8 | 385 | 460 | 5.79 | 5.08 | 8.2 |
| Z, 7.0G | 41.0 | 385 | 460 | 5.26 | 4.61 | 8.9 |

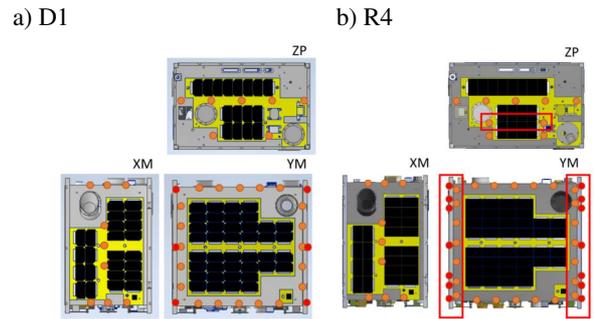

Fig. 9. Results of static load analysis: locations of maximum stress under each axis acceleration

a) D1    b) R4

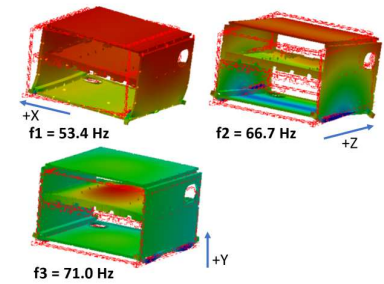

Fig. 10. Comparison of bolt connectors in D1 (a) and R4 (b)

Fig. 11. Normal modes with the lowest frequencies along each vibration axis: case using D1 bolt connectors under constraint condition C

of D1 is applied to the solid model of R4. The primary natural frequency decreases significantly from 96 Hz (R4 configuration) to 53 Hz. This result clearly indicates that the stiffness of R4 was substantially increased by the additional number of fastening bolts.

**2.4. Vibration tests and results**

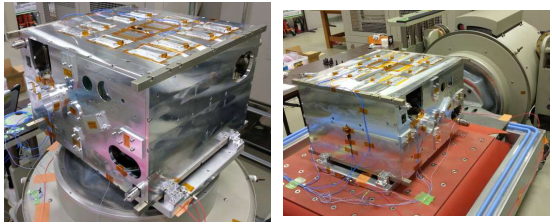

Fig. 12. MTM vibration test of R4: vertical test using a 30-mm-thick jig (left) and horizontal test (right)

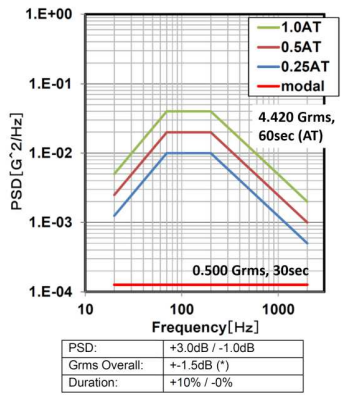

Fig. 13. Acceleration input profile for random vibration testing: frequency range of 20–2000 Hz; QT is not required

Table 5. Peak frequencies and response magnifications in the FM random vibration test

| Testbed | Axis | D1,FM modal freq. (Hz) | D1,FM AT freq. (Hz) | R4,FM modal freq. (Hz) | R4,FM AT freq. (Hz) | R4,FM PSD (G^2/Hz) | R4,FM mag. -- |
|---|---|---|---|---|---|---|---|
| Horizon | X, XP | 68.8 | 59.4 | 106 | 75.0 | 0.33 | 2.6 |
| Vertical | Y, ZP | 119 | 87.5 | 122 | 90.6 | -- | -- |
|  | Y, YP | 131 | 138 | 128 | 125 | 12.7 | 18.5 |
| Horizon | Z, ZP | 81.3 | 68.8 | 131 | 84.4 | 0.22 | 2.2 |

In the development of D1, only the flight model (FM) vibration test was conducted, requiring five days. In contrast, during the development of R4, both the Mechanical Test Model (MTM) test (two days) and the FM test (two days) were conducted. The number of days includes machine operation time as well as the installation and removal of the jigs and the satellite. The MTM consists of an assembly of the main structural panels for the FM and dummy aluminum blocks, and it does not include any electrical components.

Figure 12 shows the appearance of the R4 MTM vibration tests. Figure 13 shows the power spectral density (PSD) input level applied to the base jig, which is 4.42 Grms at the acceptance test (AT) level. The test results are summarized in Table 5.

The peak frequencies obtained in the low-level mode search and the AT vibrations are different, and the former values are adopted as the official results in the test report. Since accelerometer sensors cannot be installed inside the satellite during the FM test, three-axis accelerometers are installed at the centers of the XP, YP, and ZP panels, resulting in a total of nine measurement channels. The PSD level measured at each sensor is compared with that of the reference sensor installed on the jig surface, and the response magnification (mag.) is determined.

In the X-axis vibration test, the modal frequency of R4 increased to 106 Hz from 68.8 Hz for D1, and the response magnification at the AT level was 2.6. In the Y-axis vibration test, the YP sensor detected the local resonance of the YP panel surface, while the value measured by the ZP sensor indicated resonance of the internal CP panel. In particular, because the YP panel surface has no support at its center, the response magnification reached 18.5, which is a relatively large value and results in high stress on the solar cells. In the Z-axis vibration test, the modal frequency of R4 increased to 131 Hz from 81.3 Hz for D1, and the response magnification at the AT level was 2.2.

## 3. Countermeasures Against Bolt Loosening

### 3.1. Design and assembly processes to reduce the risk of bolt loosening

During the development of D1, only the FM vibration test was conducted due to schedule constraints. As a result, some bolts loosened at various locations during the early stages of vibration testing; however, all vibration tests were successfully completed by applying the countermeasures described below. In the development of R4, the MTM was first assembled prior to the assembly of the flight model in order to identify locations where bolt loosening could occur.

The evaluation items in the MTM test include measurement of the natural frequency of the entire satellite, measurement of response magnification at representative points (such as panel center points), and identification of loosened bolts. As described above, R4 adopted several countermeasures in advance to reduce the axial and shear loads concentrated on each bolt by increasing the number of bolts compared with D1. Nevertheless, some loosened bolts were still identified during the MTM test, and additional countermeasures were applied. As a result, all vibration tests in the FM stage were completed smoothly without any issues.

Based on these experiences, the following countermeasures were recognized as effective:

1) Bolt length is strictly managed and adjusted. The thicknesses of inserted components such as plain washers, spring washers, and heat-insulating plates are carefully controlled. It is assumed that the locking helicoil is inserted slightly deeper than the plate surface, and the bolt length should therefore be extended with an appropriate margin. In addition, the risk that a bolt cannot be fully inserted due to insufficient available thread length must be considered.

2) Increasing the fastening torque is also effective. When some bolts loosened at the specified torque value, the torque was increased by approximately 20%. If loosening still occurred, the stainless-steel bolt was replaced with a titanium bolt, and the torque value was increased by 25% relative to that used for stainless steel. Titanium bolts exhibit higher running torque than stainless-steel bolts; however, the locking portion of the helicoil may be damaged earlier, reducing the allowable number of insertion and removal cycles. Therefore, stainless-steel bolts are generally recommended when there is no significant risk of loosening.

3) The washer diameter on the bolt head side is increased.
4) The amount of vacuum-resistant grease applied to the bolt threads is reduced or eliminated.
5) Non-slip washers, such as Nord-Lock washers, are used instead of plain or spring washers.
6) The contact area between structural components to be fastened is increased. In particular, locations where only small-area spacers are used have a high risk of bolt loosening.
7) When a metal washer is placed on a resin washer for thermal insulation, both washers should have the same diameter. If the metal washer is smaller, the resin washer in contact with it may be damaged or fractured.
8) Running torque during bolt fastening is managed. If the measured torque is lower than the specified value or abnormally high, the bolt or helicoil should be replaced. When the running torque is low, it is highly likely that the bolt threads are worn, and the torque can be restored by replacing the bolt with a new one. In the MTM vibration test, bolts that were slightly worn were intentionally used, and it was confirmed that they did not loosen even when the running torque was low. These bolts were replaced with new ones during the assembly of the flight model to further reduce the risk of loosening.

### 3.2. Prediction of bolt loosening by analysis

Shear loads acting on the bolts were analyzed using two models: the R4 structural model and a quasi-D1 model. The quasi-D1 model uses the same solid elements as the R4 model, but the bolt connectors are configured to be equivalent to those of the D1 flight model. Based on previous test experience, the risk of bolt loosening tends to increase when the axial direction of a bolt is perpendicular to the vibration axis. A stainless-steel washer is inserted at the contact interface between the bolt head and the aluminum panel or resin plate surface; when these surfaces are scraped or compressed by the washer, a gap can be generated, resulting in a loss of fastening force. The conditions of the analysis model are defined as follows.
1) A static load is applied independently along each axis. In this study, the loads acting on the bolts under a 7.0 G acceleration were analyzed.
2) The beam element indices corresponding to each bolt are identified. The shear forces corresponding to each element number (in the two directions orthogonal to the bolt axial direction) are extracted from the analysis software output, and the square root of the sum of squares is calculated. The maximum shear force within each bolt group arranged in the same row is then determined. Because this analysis cannot be performed automatically using Inventor Nastran, the output options for Elemental Force and Punch format are specified. An auxiliary tool is required to automatically search and process the Punch-format output file.

Figure 14 defines the bolt group nomenclature. The figure shows the case of R4, while the number of bolts is smaller in the quasi-D1 model. All outer bolts belong to one of the defined groups. For clarity, only the minimum required number of frames is shown in the figure. For example, groups [Z1] and [Z2] exist on the ZP and ZM planes, respectively, and group [X1] appears at four locations on the XM and XP

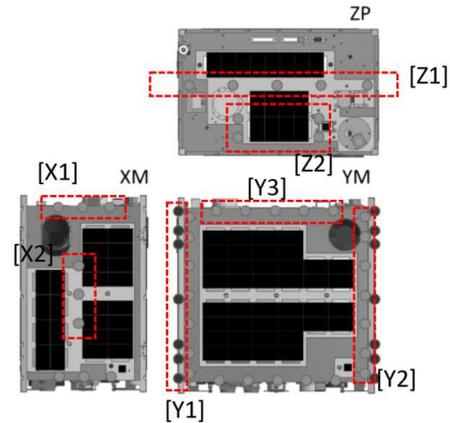

Fig. 14. Definition of bolt groups in the R4 model

Table 6. Results of shear forces in bolt (beam) elements; "L" indicates bolt loosening observed during vibration testing

| | D1 | | | | R4 | | | |
|---|---|---|---|---|---|---|---|---|
| | num. of bolts | share force (max, N) | | | num. of bolts | share force (max, N) | | |
| type | | X,7G | Y,7G | Z,7G | | X,7G | Y,7G | Z,7G |
| X1 | 12 | 311 | 230 L | 51 | 12 | 224 | 222 | 48 |
| X2 | 6 | 152 | 91 L | 413 | 6 | 152 | 81 | 410 L |
| Y1 | 12 | 707 L | 79 | 641 | 28 | 354 | 20 | 264 L |
| Y2 | 20 | 564 | 61 | 42 | 28 | 333 | 26 | 43 |
| Y3 | 20 | 311 L | 63 | 23 | 20 | 333 | 78 | 28 |
| Z1 | 10 | 484 | 116 L | 28 | 10 | 469 | 108 | 27 |
| Z2 | 4 | 48 | 353 L | 24 | 8 | 37 | 218 L | 18 |
| total | 84 | | | | 112 | | | |

planes.

Table 6 summarizes the search results for the maximum shear force in each group. The group with loose bolts during the vibration test (FM for D1 and MTM for R4) is marked with L. The vibration test order of D1 is Y-> X-> Z-> Y (retry), and R4 is Y-> X-> Z.

Table 6 summarizes the maximum shear force obtained for each bolt group. Bolt groups in which loosening occurred during vibration testing (FM for D1 and MTM for R4) are marked with "L". The vibration test sequence for D1 was Y-> X-> Z-> Y (retry), whereas that for R4 was Y-> X-> Z.

During the first vertical vibration test (Y axis) of D1, bolt loosening occurred in four groups, and the equivalent shear force under 7.0 G ranged from 91 to 353 N. At this stage, the countermeasures described earlier had not yet been applied; some bolts were too short, and some resin components were damaged by metal washers. Consequently, it is considered that bolt loosening occurred at lower load levels than in the vibration tests along the other axes.

In the horizontal vibration tests (X and Z axes) of D1, bolts in the Y1 and Y3 groups loosened during the X-axis test. This did not occur in the subsequent Z-axis test after the countermeasures had been implemented.

For R4, the number of bolts was increased during the design phase compared with D1, and the maximum shear force among all bolt groups was reduced from 707 N (X axis, Y1 group) in D1 to 469 N (X axis, Z1 group) in R4. Nevertheless, the first Y-axis vibration test using a 30-mm-thick flat jig plate caused bolt loosening in the Z2 group (218 N), and the Z-axis vibration test caused loosening in the X2 group (410 N) and the Y1 group (264 N). In both cases, the bolt material was changed to titanium and the tightening torque was increased as countermeasures.

To reduce the analysis effort, bolts with a high risk of loosening were identified using only static load analysis, without applying effective acceleration derived from Miles' equation in random vibration analysis. Comparison with the vibration test results indicates that bolts subjected to relatively large shear loads have a higher risk of loosening than others. Whether loosening occurs during vibration testing depends on the local response magnification and the material combinations at the contact interfaces. Therefore, countermeasures to reduce shear loads should first be considered during the design phase, and bolts prone to loosening should be identified in advance through MTM vibration testing. In the development of R4, this process successfully prevented any issues during the FM vibration test.

## 4. Improvement of Vertical Vibration Jig

### 4.1. Analysis of 30-mm-thick flat plate jig

In the FM vibration test of D1, a flat plate with a thickness of 30 mm was used as a jig between the satellite and the vibration testbed. Although no particular problems occurred, the R4 MTM vibration test, which used a different vibration machine, was terminated at the specified input level during vertical vibration. The catalog specifications of the machines indicated the same acceptance level; however, the machine stopped due to overload above 200 Hz. This occurred because the ratio between the limiter setting and the maximum capacity differs among test facilities. Since the natural frequency of the entire satellite was within the range of 0–200 Hz, the MTM test was conducted only within this frequency range. In contrast, for the FM vibration test, the jig needed to be redesigned for vertical vibration, and the natural frequencies of the satellite and the jig plate needed to be sufficiently separated.

As shown in Fig. 15, the natural frequency was 197 Hz when a 50-kg point mass element was rigidly fastened to the flat plate. This result is consistent with the actual test observation that a large resonance occurred around 200 Hz.

### 4.2. Design of a higher-stiffness vertical vibration jig

To increase the natural frequency when the satellite is fixed to the jig, it is necessary to increase the stiffness of the jig itself. Increasing the plate thickness can improve stiffness; however, the increased mass has a negative effect on stiffness. As a result, stiffness does not increase efficiently, and the jig mass can become excessively large.

In general, the maximum mass that can be handled by a single person is approximately 25 kg. This limitation also applies when the jig is transported by a courier and handled manually by a single delivery driver. Therefore, the target jig mass was limited to 22.5 kg, corresponding to 90% of 25 kg.

Although the jig thickness was increased, it was necessary to reduce the weight by removing material from regions that do not significantly contribute to stiffness. First, the central portion was cut out while leaving the bottom plate and the outer frame. Since the bottom fixing interface has a circular shape with an outer diameter of 30 cm, the bottom side was tapered to reduce weight without degrading stiffness. As a result, the shape shown in Fig. 16 was designed, and the natural frequency increased to approximately 410 Hz when combined with a 50-kg mass.

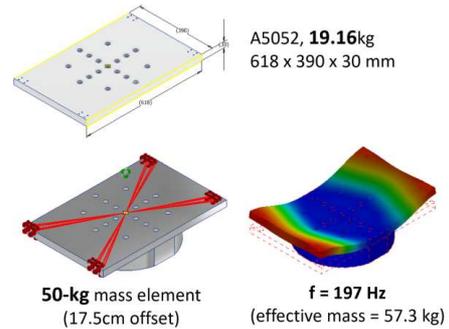

Fig. 15. Normal mode analysis of a 30-mm-thick jig with a 50-kg mass element: first mode at 197 Hz

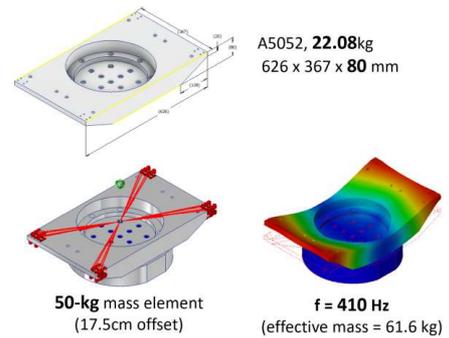

Fig. 16. Normal mode analysis of an 80-mm-thick jig with a 50-kg mass element: first mode at 410 Hz

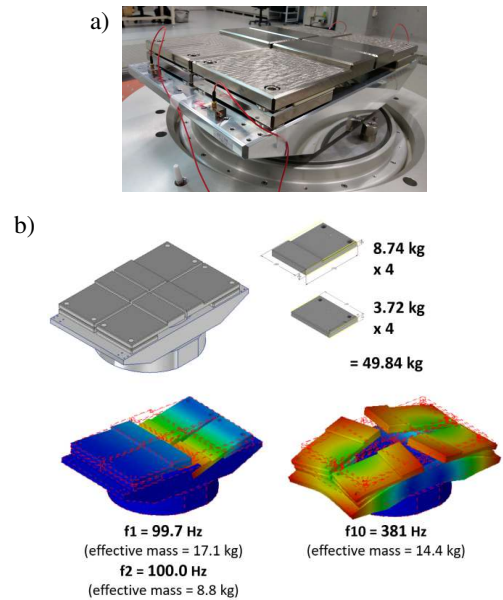

Fig. 17. Vibration test using a 50-kg dummy mass with an 80-mm-thick jig: (a) setup appearance, (b) normal mode analysis result (381 Hz)

### 4.3. Design of a 50-kg dummy mass for evaluation of the vertical vibration testbed

Even when the model and performance of vibration machines are identical, the maximum acceptable vibration force may be limited by the operational policy of each facility. In addition, the control method of the testbed may differ among facilities, and the number of control sensors can be selected from one, two, or four; however, the available number of sensors is constrained depending on the facility.

For the FM vibration test of R4, vertical vibration tests were conducted as trials at three test sites (one day each) using the newly designed vertical vibration jig and a dummy mass equivalent to 50 kg. Since the jig and dummy mass were designed to be transported in packages weighing less than 25 kg each, the tests could be carried out efficiently by changing test locations.

Figure 17 shows the shape of the dummy mass and the actual installation configuration. By dividing the 50-kg mass into four parts (12.5 kg each) and further splitting each part into two steel plates, the natural frequency when coupled with the jig was at least 100 Hz (with an effective mass of 25.9 kg in total for the two modes). The resonance with the jig occurred at 381 Hz with an effective mass of 14.4 kg. This configuration reproduces the natural frequency characteristics of the satellite (96–97 Hz, 23.3 kg), and it can therefore be concluded that the FM vibration test can be safely conducted once the dummy mass can be vibrated without any issues.

The control method of the vibration machine is also important. Two approaches are possible: allowing overshoot to reach the target Grms in the shortest time, or gradually approaching the target Grms over a longer duration. In this study, the former control strategy was adopted for the vertical vibration test to reduce overall damage to the satellite, and the average signal from four sensors was used as the control reference.

### 5. Conclusion

In this paper, valuable experiences in structural design and vibration test evaluation for 50-cm-class microsatellites deployed from the ISS were presented. Two satellite development opportunities were addressed, and based on the issues encountered during the first development, new methods were applied and their effectiveness was demonstrated in the second satellite. Section 2 presented the design methodology, analysis results, and test evaluation results. Section 3 described a prediction method for bolt loosening based on countermeasures and analysis, which addresses one of the most common issues encountered in satellite vibration testing. It is important to conduct MTM testing with dummy components in advance to avoid issues during the final FM test. Section 4 presented examples of problems associated with vertical vibration jigs and the corresponding improvements, highlighting the importance of preliminary evaluations at multiple test sites using dummy masses.

In the development of R4, approximately 1.5 months were required from FM assembly to completion of the vibration tests, and about one month later, the satellite was handed over to the launch organization. Shortening the development period is important for reducing development costs. In addition, it allows more time to be allocated to onboard software development, including refinement of the attitude control system. It is hoped that the experiences described in this paper will contribute to the success of future satellite projects.

### bibliographic note

### References

1) Junichi Kurihara, Toshinori Kuwahara, Shinya Fujita, Yuji Sato, Kosuke Hanyu, Morokot Sakal, Yu Murata, Hannah Tomio, Yukihiro Takahashi, and Wing Huen Ip, "A High Spatial Resolution Multispectral Sensor on the RISESAT Microsatellite", *Trans. JSASS Aerospace Technology Japan*, Vol. 18, No 5, pp.186 191, 2020.

2) Kurihara, J; Takahashi, Y; Sakamoto, Y; Kuwahara, T; Yoshida, K, "HPT: A High Spatial Resolution Multispectral Sensor for Microsatellite Remote Sensing," *Sensors*, 2018, 18(2), 619, Feb. 2018, DOI: 10.3390/s18020619.

3) Yuji Sakamoto, Toshinori Kuwahara, Shinya Fujita, and Kazuya Yoshida, "International Cooperation of Education for Young Engineers and Scientists by the Development and Flight Operations for Philippine DIWATA Microsatellites," *Proc. 2019 JSEE Annual Conference & AEESEAP Workshop*, 2019, W-08.

4) Shinya FUJITA, Yuji SATO, Toshinori KUWAHARA, Yuji SAKAMOTO, Kazuya YOSHIDA, "Development and Ground Evaluation of Ground-Target Tracking Control of Microsatellite RISESAT", *TRANSACTIONS OF THE JAPAN SOCIETY FOR AERONAUTICAL AND SPACE SCIENCES, AEROSPACE TECHNOLOGY JAPAN*, 2019 Volume 17 Issue 2 Pages 120-126.

5) Shinya Fujita; Yuji Sato; Toshinori Kuwahara; Yuji Sakamoto; Kazuya Yoshida, "Attitude Maneuvering Sequence Design of High-Precision Ground Target Tracking Control for Multispectral Earth Observations", *2019 IEEE/SICE International Symposium on System Integration (SII)*.

6) Yuji Sato; Shinya Fujita; Toshinori Kuwahara; Hiroto Katagiri; Yuji Sakamoto; Kazuya Yoshida, "Improvement and verification of satellite dynamics simulator based on flight data analysis", *2017 IEEE/SICE International Symposium on System Integration (SII)*.